\documentclass{emulateapj}
\usepackage{natbib}
\begin{document}

\title{A Black Hole Mass Determination for the Compact Galaxy Mrk
  1216}

\author{Jonelle L. Walsh$^{1}$, Remco C.~E. van den Bosch$^{2}$, Karl
  Gebhardt$^{3}$, Ak{\i}n Y{\i}ld{\i}r{\i}m$^{2,4}$, Kayhan
  G\"{u}ltekin$^{5}$, Bernd Husemann$^{2,6}$, {\sc and} Douglas
  O. Richstone$^{5}$}

\affil{$^1$ George P. and Cynthia Woods Mitchell Institute for
  Fundamental Physics and Astronomy, Department of Physics and
  Astronomy, Texas A\&M University, 4242 TAMU, College Station, TX
  77843, USA; walsh@physics.tamu.edu \\
  $^2$ Max-Planck-Institut f\"{u}r Astronomie, K\"{o}nigstuhl 17,
  D-69117 Heidelberg, Germany\\
  $^3$ Department of Astronomy, The University of Texas at Austin,
  2515 Speedway, Stop C1400, Austin, TX 78712, USA \\
  $^4$ Max-Planck-Institut f\"{u}r Astrophysik,
  Karl-Schwarzschild-Str.~1, 85741 Garching, Germany \\
  $^5$ Department of Astronomy, University of Michigan, 1085
  S.~University Ave., Ann Arbor, MI 48109, USA \\
  $^6$ European Southern Observatory, Karl-Schwarzschild-Str.~2, 85748
  Garching, Germany}

\begin{abstract}

  Mrk 1216 is a nearby, early-type galaxy with a small effective
  radius of 2.8 kpc and a large stellar velocity dispersion of 308 km
  s$^{-1}$ for its $K$-band luminosity of $1.4\times10^{11}\
  L_\odot$. Using integral-field spectroscopy assisted by adaptive
  optics from Gemini North, we measure spatially resolved stellar
  kinematics within $\sim$450 pc of the galaxy nucleus. The galaxy
  exhibits regular rotation with velocities of $\pm 180$ km s$^{-1}$
  and a sharply peaked velocity dispersion profile that reaches 425 km
  s$^{-1}$ at the center. We fit axisymmetric, orbit-based dynamical
  models to the combination of these high angular resolution
  kinematics, large-scale kinematics extending to roughly three
  effective radii, and \emph{Hubble Space Telescope} imaging,
  resulting in a constraint of the mass of the central black hole in
  Mrk 1216. After exploring several possible sources of systematics
  that commonly affect stellar-dynamical black hole mass measurements,
  we find a black hole mass of $M_\mathrm{BH} = (4.9\pm1.7)\times10^9\
  M_\odot$ and a $H$-band stellar mass-to-light ratio of $\Upsilon_H =
  1.3\pm0.4\ \Upsilon_\odot$ (1$\sigma$ uncertainties). Mrk 1216 is
  consistent with the local black hole mass -- stellar velocity
  dispersion relation, but is a factor of $\sim$$5-10$ larger than
  expectations from the black hole mass -- bulge luminosity and black
  hole mass -- bulge mass correlations when conservatively using the
  galaxy's total luminosity or stellar mass. This behavior is quite
  similar to the extensively studied compact galaxy NGC
  1277. Resembling the $z \sim 2$ quiescent galaxies, Mrk 1216 may be
  a passively evolved descendant, and perhaps reflects a previous era
  when galaxies contained over-massive black holes relative to their
  bulge luminosities/masses, and the growth of host galaxies had yet
  to catch up.

\end{abstract}


\section{Introduction}
\label{sec:intro}

Our understanding of the connection between supermassive black holes
and their host galaxies is anchored by $\sim$100 dynamical black hole
mass ($M_\mathrm{BH}$) measurements that have been made over the past
two decades \citep[e.g.,][and references therein]{Kormendy_Ho_2013,
  vandenBosch_2016}. Strong correlations have emerged between
$M_\mathrm{BH}$ and large-scale galaxy properties, like the bulge
luminosity ($L_\mathrm{bul}$; e.g., \citealt{Kormendy_Richstone_1995,
  Marconi_Hunt_2003, Kormendy_Ho_2013}) or mass ($M_\mathrm{bul}$;
e.g., \citealt{Haring_Rix_2004, Sani_2011, McConnell_Ma_2013}), and
the stellar velocity dispersion ($\sigma_\star$; e.g.,
\citealt{Ferrarese_Merritt_2000, Gebhardt_2000, Gultekin_2009}). These
relations are connected to each other, and the search for the most
fundamental one is still ongoing \citep{Beifiori_2012, Saglia_2016,
  vandenBosch_2016}. The empirical relationships imply that black
holes are key components of galaxies and regulate galaxy properties
via feedback mechanisms \citep{Silk_Rees_1998, Fabian_1999}, although
a non-causal origin in which black holes do not actively shape their
host galaxies is also possible \citep{Peng_2007,
  Jahnke_Maccio_2011}. Establishing the exact role of black holes in
galaxy evolution and accurately inferring the black hole mass function
requires increasing the number of $M_\mathrm{BH}$ measurements,
specifically targeting galaxies with diverse properties that have
experienced varied growth channels.

As we begin to examine a broader range of black hole masses and hosts,
galaxies with different structural properties show surprises in the
scaling relations. Recent progress detecting high-mass black holes in
Brightest Cluster Galaxies (BCGs) and other large early-type galaxies
hint that these objects may be positive outliers from $M_\mathrm{BH} -
\sigma_\star$ and $M_\mathrm{BH} - L_\mathrm{bul}$
\citep[e.g.,][]{McConnell_2012, Rusli_2013, Thomas_2016}, but there
are still too few measurements to firmly characterize the scaling
relations at $M_\mathrm{BH} \gtrsim 10^9\ M_\odot$. The uncertainties
are equally severe at the opposite end, where spiral galaxies with
low-mass black holes ($M_\mathrm{BH} \lesssim 10^7\ M_\odot$) measured
from water megamaser disks exhibit substantial scatter below the
global black hole -- host galaxy relations
\citep[e.g.,][]{Greene_2010, Lasker_2016, Greene_2016}. There are also
new observations of compact galaxies, whose black holes are a
remarkably large fraction of the galaxy's stellar mass
\citep[e.g.,][]{Seth_2014, Walsh_2015, Walsh_2016, Saglia_2016}.

NGC 1277 and NGC 1271 are two such compact galaxies, with NGC 1277
being widely studied over the last few years
\citep[e.g.,][]{vandenBosch_2012, Emsellem_2013, Walsh_2016,
  Scharwachter_2016, Graham_2016a}. Both were originally discovered by
the HET Massive Galaxy Survey \citep{vandenBosch_2015} and share
considerable similarities with the $z \sim 2$ quiescent galaxies
(e.g., \citealt{Trujillo_2014, FerreMateu_2015, Yildirim_2015};
Y{\i}ld{\i}r{\i}m {et~al.} 2016b, in prep). Like the $z \sim 2$ red
nuggets, NGC 1277 and NGC 1271 have small effective radii of $R_e \sim
1-2$ kpc, stellar masses of $M_\star \sim 10^{11}\ M_\odot$, and
stellar mass surface density profiles that are elevated at the center
and drop off steeply at larger radii compared to low-redshift
early-type galaxies. In addition, NGC 1277 and NGC 1271 are rotating,
consistent with the disk-like flattened structures of the $z \sim 2$
red nuggets \citep{vanderWel_2011}, and have uniformly old stellar
populations (ages of $\sim$10 Gyr) extending out to several $R_e$. The
red nuggets are thought to grow in size and moderately in mass through
mergers to produce the present-day massive galaxies
\citep[e.g.,][]{vanDokkum_2010}, with a small fraction experiencing
passive evolution since $z \sim 2$ \citep[e.g.,][]{Trujillo_2009,
  Wellons_2016}. NGC 1277 and NGC 1271 appear to be such relics of the
red nuggets, and could provide the unique opportunity to gain insight
into the black holes at earlier epochs.

Of the current galaxies with dynamical $M_\mathrm{BH}$ measurements,
NGC 1277 and NGC 1271 are most similar to NGC 4342
\citep{Cretton_vandenBosch_1999} and NGC 1332. All are nearby galaxies
that are flattened and rotating, with small effective radii and large
stellar velocity dispersions for their luminosities. They contain
black holes that are more massive than the predictions from
$M_\mathrm{BH} - L_\mathrm{bul}$, yet are consistent with
$M_\mathrm{BH} - \sigma_\star$. The magnitude of the offset from
$M_\mathrm{BH} - L_\mathrm{bul}$ depends on the adopted bulge
luminosity \citep[e.g.,][]{Walsh_2015, Walsh_2016, Graham_2016a,
  Graham_2016b}, although we note that NGC 1332 may very well be
consistent with the black hole relations given the uncertainties
associated with both $M_\mathrm{BH}$ \citep{Rusli_2011, Barth_2016a,
  Barth_2016b} and the bulge component \citep{Rusli_2011,
  Kormendy_Ho_2013, Savorgnan_Graham_2016, Saglia_2016}. Clearly,
additional mass measurements of black holes in NGC 1277-like galaxies
are needed. A significant sample of local analogs to the $z \sim 2$
red nuggets that are also positive outliers from $M_\mathrm{BH} -
L_\mathrm{bul}$ would suggest that this black hole scaling relation
did not apply at earlier times, galaxies instead harbored over-massive
black holes, and subsequent galaxy growth had yet to occur.

Beyond the connections to the $z \sim 2$ red nuggets, the compact
galaxies are interesting because they occupy the sparsely populated
upper-end of the $M_\mathrm{BH}$ relationships, and their properties
are quite distinct from the BCGs and giant ellipticals that are
expected to house the most massive black holes in the
Universe. Currently, at the high-mass end, the differing behaviors of
$M_\mathrm{BH} - \sigma_\star$ and $M_\mathrm{BH} - L_\mathrm{bul}$,
and the poorly characterized scatter in $M_\mathrm{BH}$ for fixed
$\sigma_\star$ or $L_\mathrm{bul}$, lead to strongly divergent
predictions of the black hole mass function \citep{Lauer_2007a}. This
in turn affects inferences about black hole growth histories and
constraints of the mean radiative efficiency of black hole accretion,
the duty cycle of active galactic nuclei, and the redshift evolution
of the scaling relations \citep[e.g.,][]{Marconi_2004, Lauer_2007b,
  Shankar_2009, Robertson_2006}.

In this paper, we examine Mrk 1216, an early-type, compact
($R_e = 2.8$ kpc, $M_\star = 1.6\times10^{11}\ M_\odot$),
high-dispersion ($\sigma_\star = 308$ km s$^{-1}$) galaxy found
through the HET Massive Galaxy Survey. \cite{Yildirim_2015} presented
wide-field integral-field spectroscopy of Mrk 1216, and constructed
axisymmetric Schwarschild models in order to learn about the galaxy's
dynamical stellar mass-to-light ratio and dark matter halo. The
spatial resolution, however, was insufficient to pin down the black
hole mass, and \cite{Yildirim_2015} set an upper-limit of
$M_\mathrm{BH} < 1\times10^{10}\ M_\odot$. Here, we use Gemini North
observations assisted by adaptive optics (AO) to resolve the region
where the black hole dominates the gravitational potential (the black
hole sphere of influence;
$r_\mathrm{sphere} = G M_\mathrm{BH}/\sigma_\star^2$), thereby
obtaining a secure stellar-dynamical $M_\mathrm{BH}$ measurement. We
assume a distance of 94 Mpc to Mrk 1216. This is the same distance
adopted by \cite{Yildirim_2015} and is the Virgo + Great Attractor +
Shapley Supercluster Infall value \citep{Mould_2000} for $H_0 = 70.5$
km s$^{-1}$ Mpc$^{-1}$, $\Omega_M = 0.27$ and $\Omega_\Lambda =
0.73$. At this distance, 1\arcsec\ corresponds to 456 pc.

The paper is structured as follows. We review the imaging observations
and the luminous mass model in Section \ref{sec:hst_img}. In Sections
\ref{sec:nifs_obs_measurements} and \ref{sec:largescale_spec}, we
describe the high angular resolution and the large-scale spectroscopic
observations, the measured stellar kinematics, and the point-spread
function (PSF) characterizations. An overview of the stellar-dynamical
models and the results from those models, including an examination of
the black hole mass error budget, is provided in Section
\ref{sec:stellardyn_models}. We study the galaxy's orbital structure
in \ref{sec:orbitstructure}, and discuss the location of Mrk 1216 on
the black hole mass -- host galaxy relationships, as well as the
implications, in Sections \ref{sec:bhgalrels} and
\ref{sec:discussion}. Concluding remarks are provided in Section
\ref{sec:conclusion}.

\section{HST Imaging}
\label{sec:hst_img}

We obtained an \emph{HST} Wide-Field Camera 3 (WFC3) $F160W$ image of
Mrk 1216. The WFC3/IR observations were executed under program
GO-13050, and included dithered full array images and brief subarray
exposures with a total integration time of 1354 s. The data were
reduced, and the flattened, calibrated images were corrected for
geometric distortions, cleaned, and combined using AstroDrizzle
\citep{Gonzaga_2012} to produce a super-sampled image with a scale of
0\farcs06 pixel$^{-1}$. After masking the foreground stars, we
described the galaxy's stellar surface brightness distribution as the
sum of two-dimensional (2D) Gaussians. Such a Multi-Gaussian Expansion
\citep[MGE;][]{Monnet_1992, Emsellem_1994} is able to recover the
surface brightness profiles of realistic multi-component galaxies
while also allowing for the intrinsic luminosity density to be
determined through an analytical deprojection. During the MGE fit, we
took into account the WFC3 PSF from \cite{vanderWel_2012}. The PSF was
generated with TinyTim \citep{Krist_Hook_2004} for the $F160W$ filter
and a G2~V star at the center of the WFC3 detector, then drizzled to
produce the same scale as our final Mrk 1216 image.

The Mrk 1216 MGE is composed of 10 Gaussians with dispersions,
measured along the major axis, of 0\farcs09$-$29\farcs76, and
projected axis ratios between 0.52 and 0.99. The components have the
same center, and a position angle of 70.2$^\circ$ east of north. The
final parameter values, after correction for Galactic extinction using
the \cite{Schlafly_Finkbeiner_2011} WFC3 $F160W$ value of 0.017 mag
and assuming an $H$-band absolute solar magnitude of 3.32
\citep{Binney_Merrifield_1998}, are given in \cite{Yildirim_2015}. The
MGE fits the \emph{HST} image very well within the central
$\sim$40\arcsec, and allows us to accurately infer the stellar
gravitational potential. We refer the reader to \cite{Yildirim_2015}
for additional details regarding the imaging observations, data
reduction, and construction of the MGE model.

\section{NIFS Observations and Measurements}
\label{sec:nifs_obs_measurements}

In addition to the luminous mass model, stellar kinematics on scales
comparable to the black hole sphere of influence are crucial inputs
into the dynamical models. We therefore observed Mrk 1216 with the
Near-infrared Integral Field Spectrometer
\citep[NIFS;][]{McGregor_2003} aided by the ALTtitude conjugate
Adaptive optics for the InfraRed \citep{Herriot_2000, Boccas_2006}
system on Gemini North. The observations were taken on 21 Dec 2013
under program GN-2013A-Q-1 in laser guide star (LGS) mode using an
$R=13.7$ mag star located 22\arcsec\ from the galaxy nucleus as the
tip-tilt reference. Four blocks of consecutive Object-Sky-Object
sequences with 600 s exposures were recorded. The observations were
acquired with the $H+K$ filter and the $K$ grating centered on $2.2$
$\mu$m. We observed the tip-tilt star once during the night for an
estimate of the PSF and an A0~V star for telluric correction. The
normal baseline calibrations consisting of dark frames, flat fields,
Argon/Xeon arc lamp exposures, and a Ronchi mask (to establish the
spatial rectification) were taken as well.

We processed the NIFS data using PyRAF \footnote[1]{PyRAF is a product
  of the Space Telescope Science Institute, which is operated by AURA
  for NASA} and the Gemini data reduction package version 1.11,
following the steps in the NIFS example
scripts\footnote[2]{http://www.gemini.edu/sciops/instruments/nifs/data-format-and-reduction}
for calibration, telluric, and science exposures. For the galaxy, the
basic procedure consisted of preparing the raw images for processing
within the NIFS data reduction package, subtracting sky frames from
adjacent object exposures, flat fielding, removing bad pixels and
cosmic rays, wavelength calibration, and spatial rectification. We
corrected for telluric features using an A0~V star, whose spectrum had
been divided by a blackbody with a temperature of 9480 K after
interpolating over the Br$\gamma$ absorption line. We then assembled
data cubes having $x$ and $y$ spatial dimensions, each with a scale of
0\farcs05 pixel$^{-1}$, and a wavelength axis. We determined the
spatial offsets between individual galaxy cubes by summing over the
wavelength dimension and cross-correlating the images. These eight
cubes, corresponding to a total of 1.3 hours on-source, were aligned
and combined to produce the final Mrk 1216 data cube. We followed
similar steps to reduce the NIFS observation of the PSF star.

\subsection{Stellar Kinematics}
\label{subsec:nifs_kinematics}

From the reduced Mrk 1216 data cube, we measured the stellar
kinematics as a function of spatial location. Specifically, we
extracted the line-of-sight velocity distribution (LOSVD),
parameterized by the first four Gauss-Hermite moments, in 67 Voronoi
spatial bins \citep{Cappellari_Copin_2003} using the penalized pixel
fitting (pPXF) code of \cite{Cappellari_Emsellem_2004}. The spatial
bins were chosen so that the galaxy spectra had a signal-to-noise
ratio (S/N) $\gtrsim$ 40, where the S/N was measured as the median
flux divided by the standard deviation of the pPXF model
residuals. Such a high S/N spectrum is required in order to measure
the LOSVD's deviation from a Gaussian
\citep[e.g.,][]{vanderMarel_Franx_1993, Bender_1994}.

\begin{figure}
\begin{center}
\epsscale{1.0}
\plotone{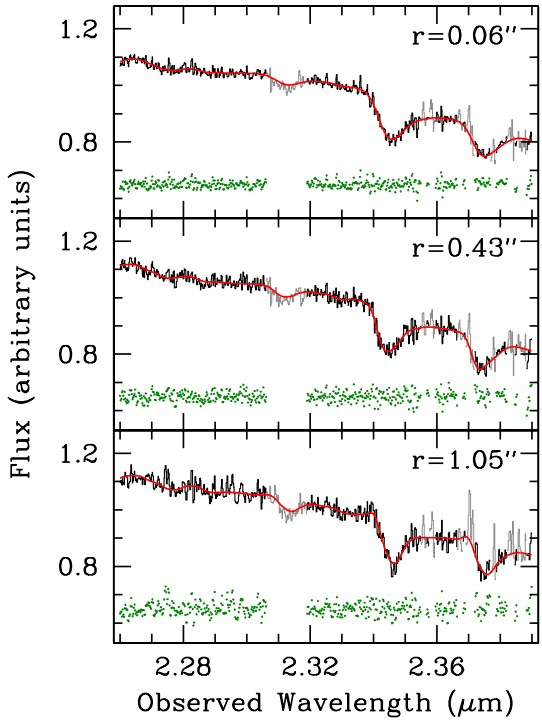}
\caption{Example fits with pPXF to the observed Mrk 1216 spectra
  located at the nucleus (top), at an intermediate location (middle),
  and in one of the outermost spatial bins (bottom). The red line
  shows the optimal stellar template convolved with the LOSVD, whose
  shape is further adjusted with an additive constant and a
  first-degree multiplicative Legendre polynomial. Several wavelength
  regions shown in gray were masked during fit, including the Ca I
  absorption line and artifacts from imperfect data reduction. The
  green dots are the model residuals that have been shifted upward by
  a constant, arbitrary amount. \label{fig:nifs_specfit}}
\end{center}
\end{figure}

We provided pPXF with a velocity template library composed of 12 stars
(K0$-$M5 giant stars and two late-type supergiants), which were
observed with NIFS in the $K$ band. The stars are a subset of those
presented in \cite{Winge_2009}, but we have reduced the data ourselves
\citep{Walsh_2016}, starting with the raw frames and their calibration
files retrieved from the Gemini Science Archive. During the fit with
pPXF, we corrected for slight differences in the continuum shape and
equivalent width between the LOSVD-convolved template stars and the
observed galaxy spectra via an additive constant and a multiplicative
Legendre polynomial of degree 1. The LOSVD was largely constrained by
the strong $^{12}$CO(2$-$0) and $^{12}$CO(3$-$1) bandheads, which were
contained within the $2.26-2.39$ $\mu$m fitting region. We masked the
\ion{Ca}{1} absorption line because our template library does not
include cool dwarf stars \citep[e.g.,][]{Krajnovic_2009}, and further
excluded a few artificial features, likely the result of imperfect sky
subtraction or telluric correction. Example fits with pPXF to the
observed galaxy spectra located at the nucleus, in an intermediate
region, and in one of the outermost spatial bins are given in Figure
\ref{fig:nifs_specfit}.

After an initial fit to the galaxy spectrum in each spatial bin, we
ran a Monte Carlo simulation with 100 iterations. During each
realization, we generated a synthetic spectrum by taking the best-fit
model and adding random Gaussian noise based upon the standard
deviation of the model residuals. We re-fit the spectrum using pPXF
with the penalization turned off. From the resulting distribution for
each Gauss-Hermite moment, we took the mean to be the kinematic value
and the standard deviation to be the 1$\sigma$ uncertainty. Finally,
we point-symmetrize the kinematics using the method described in
\cite{vandenBosch_deZeeuw_2010}, which also removes the systematic
offsets in the odd Gauss-Hermite moments.

\begin{figure*}
\begin{center}
\epsscale{0.85}
\plotone{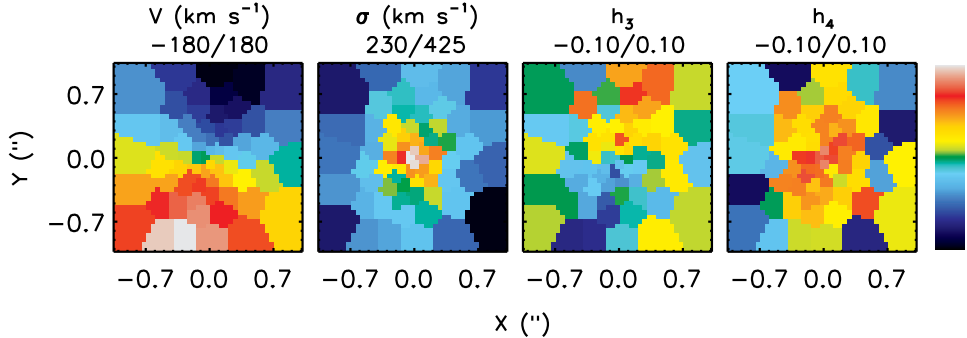}
\caption{The Mrk 1216 stellar kinematics from NIFS assisted by LGS
  AO. The measurements are shown on the scale given by the color bar
  to the right, and the range of values is provided at the top of the
  $V$, $\sigma$, $h_3$, and $h_4$ maps. Mrk 1216 exhibits regular
  rotation, a central rise in the stellar velocity dispersion, a
  $h_3 - V$ anti-correlation, and a slight peak in $h_4$ at the
  nucleus. The blueshifted velocities correspond to the southwest side
  of the galaxy. \label{fig:nifs_maps}}
\end{center}
\end{figure*}

The resulting radial velocity ($V$) map shows that Mrk 1216 is
rotating, such that the southwest side of the galaxy is blueshifted
and the northeast side is redshifted with values of $\pm$180 km
s$^{-1}$. The velocity dispersion ($\sigma$) rises from 230 km
s$^{-1}$ at a projected radius of $\sim$1\arcsec\ to 425 km s$^{-1}$
at the nucleus. The third Gauss-Hermite moment ($h_3$), or skewness,
falls between $\pm$0.07, and we observe a $h_3 - V$ anti-correlation,
which is a common for rotating, axisymmetric systems
\citep[e.g.,][]{Fisher_1997}. The map of the fourth Gauss-Hermite
moment ($h_4$), or the kurtosis, has a slight peak at the nucleus to a
value of 0.08. The kinematics have median errors of 15 km s$^{-1}$, 18
km s$^{-1}$, 0.04, and 0.04 for $V$, $\sigma$, $h_3$, and $h_4$,
respectively. Table \ref{tab:nifskin} provides the extracted
Gauss-Hermite moments for each NIFS spatial bin, and the 2D velocity
fields are shown in Figure \ref{fig:nifs_maps}.

\begin{deluxetable*}{rrrrrrrrrr}
\tabletypesize{\scriptsize}
\tablewidth{0pt}
\tablecaption{NIFS Kinematics \label{tab:nifskin}}
\tablehead{
\colhead{$x$} &
\colhead{$y$}&
\colhead{$V$} &
\colhead{$\Delta V$} &
\colhead{$\sigma$} &
\colhead{$\Delta \sigma$} &
\colhead{$h_3$} &
\colhead{$\Delta h_3$} &
\colhead{$h_4$} &
\colhead{$\Delta h_4$} \\
\colhead{(\arcsec)} &
\colhead{(\arcsec)} &
\colhead{(km s$^{-1}$)} &
\colhead{(km s$^{-1}$)} &
\colhead{(km s$^{-1}$)} &
\colhead{(km s$^{-1}$)} &
\colhead{(7)} &
\colhead{(8)} &
\colhead{(9)} &
\colhead{(10)}
}

\startdata

  -0.055  &   0.003  &    -1.205  &  12.708  &   425.376  &  17.927  &  -0.010  &   0.024  &   0.075  &  0.031  \\
   0.012  &  -0.069  &    54.401  &  14.221  &   424.049  &  20.599  &  -0.023  &   0.028  &   0.082  &  0.034  \\
   0.013  &   0.079  &   -52.747  &  14.410  &   420.777  &  20.519  &   0.024  &   0.028  &   0.083  &  0.033  \\
   0.087  &  -0.012  &    17.161  &  13.271  &   415.750  &  18.381  &  -0.000  &   0.026  &   0.072  &  0.034  \\
   0.107  &   0.092  &   -46.839  &  13.620  &   369.604  &  17.681  &   0.031  &   0.027  &   0.063  &  0.033

\enddata

\tablecomments{The first two columns provide the $x$ and $y$ locations
  of the Voronoi bin generators, while the remaining columns present
  the point-symmetrized NIFS kinematics and their uncertainties. The
  position angle is 283.94$^\circ$, defined counterclockwise from the
  galaxy's major axis to $x$. This table is available in its entirety
  in machine-readable form.}

\end{deluxetable*}

We also examined the robustness of the stellar kinematics by changing
how the measurements were made with pPXF. We modified the fitting
region to be longer ($2.26-2.43$ $\mu$m) and shorter ($2.26-2.37$
$\mu$m), included the Ca I absorption line in the fit, required that
the relative mix of template stars remain the same between spatial
bins, and tested different combinations of degree $0-2$ multiplicative
polynomials along with no additive component, an additive constant,
and first-degree additive Legendre polynomial. We found the kinematics
were consistent within 2$\sigma$ compared to our default fitting
approach. We note that there were a number of bins whose kinematics
were inconsistent at the 1$\sigma$ level, with the number of
discrepant bins ranging from $0-15$ depending on the fitting method
being adopted. In Section \ref{subsec:moremodels}, we test the effect
on $M_\mathrm{BH}$ if the kinematics from an alternative fitting
approach in which a multiplicative degree 2 polynomial and no additive
term is used instead.

\subsection{PSF Model}
\label{subsec:nifs_psf}

We described the NIFS PSF as the sum of two concentric, circular
Gaussians. Following past work \citep[e.g.,][]{Krajnovic_2009,
  Seth_2014}, the MGE in Section \ref{sec:hst_img} was convolved with
the NIFS PSF and compared to the Mrk 1216 data cube, after collapsing
along the wavelength axis. This resulted in best-fit values of
0\farcs07 and 0\farcs36 for the dispersions, and 0.56 and 0.44 for the
weights of the inner and and outer Gaussian components,
respectively. The method further allowed us to determine the center of
the NIFS aperture. Attempts to fit a three-Gaussian PSF model produced
a negligible component that contributed 1\% to the total flux. Our
characterization of the NIFS PSF is consistent with expectations of
the Gemini ALTAIR system \citep[e.g.,][]{Gebhardt_2011, Onken_2014,
  Drehmer_2015}.

We also measured the PSF using an LGS AO NIFS observation of the
galaxy's tip-tilt star. With the 2D image decomposition package Galfit
\citep{Peng_2010}, we found that the sum of three circular Gaussians
fit the collapsed NIFS data cube of the star well. The Gaussians have
dispersions of 0\farcs07, 0\farcs14, and 0\farcs37 with weights of
0.39, 0.21, and 0.40. Due to the temporal variability of the AO
correction, and because the observations of the star were conducted
on-axis in contrast to the off-axis observations of the galaxy, we
view this second PSF determination as a rough estimate. Nevertheless,
we use this result to test how sensitive the inferred $M_\mathrm{BH}$
is to the assumed PSF in Section \ref{subsec:model_results}.

\section{Large-Scale Spectroscopy}
\label{sec:largescale_spec}

The NIFS kinematics are complemented by large-scale spectroscopic
observations that provide important constraints on the stellar
mass-to-light ratio and the orbital distribution
\citep[e.g.,][]{Shapiro_2006}. The large-scale spectra were obtained
with the Potsdam Multi Aperture Spectrograph
\citep[PMAS;][]{Roth_2005} in the Pmas fiber PAcK
\citep[PPAK;][]{Verheijen_2004, Kelz_2006} mode from the 3.5 m
telescope at Calar Alto Observatory, and from the Marcario
Low-Resolution Spectrograph \citep[LRS;][]{Hill_1998} on the
Hobby-Eberly Telescope at McDonald Observatory. \cite{Yildirim_2015}
and \cite{vandenBosch_2015} presented the PPAK and HET data, but we
provide a brief summary below.

The PPAK integration time was 1.5 hours on-source, with two 900 s
exposures taken at three dither positions to fully sample the 331
2\farcs7-wide science fibers. We acquired the data on 5 Dec 2011 with
the medium resolution V1200 grating, covering $3650-4620$ \AA\ with a
spectral resolving power of $R$$\sim$1650 at 4000 \AA. Data reduction
followed the approach of the Calar Alto Legacy Integral Field Area
Survey \citep{Sanchez_2012, Husemann_2013}. We extracted $V$,
$\sigma$, $h_3$, and $h_4$ in 41 Voronoi spatial bins using pPXF, the
Indo-U.S. Library of Coud\'{e} Feed Stellar Spectra
\citep{Valdes_2004}, and an additive Legendre polynomial of degree
15. As a final step, we point-symmetrized the stellar kinematics. The
PSF was reconstructed by comparing the collapsed PPAK data cube to the
Mrk 1216 MGE. The PSF has an inner Gaussian component with a
dispersion of 1\farcs24 that contributes 77\% to the total flux, while
the second Gaussian component has a weight of 23\% and a dispersion of
3\farcs72.

Moreover, we have a single 900 s exposure of Mrk 1216 taken with the
HET/LRS 2\arcsec-wide slit aligned with the galaxy major axis. The g2
grating and $2\times2$ binning provided coverage of $4200-7400$ \AA\
and an instrumental dispersion of 180 km s$^{-1}$. After the initial
data processing, we constructed 21 spatial bins and measured the
stellar kinematics with pPXF and the MILES template library
\citep{SanchezBlazquez_2006, FalconBarroso_2011}. Measurements of $V$
and $\sigma$ were made in each of the spatial bins, with $h_3$ and
$h_4$ being extracted from the inner 10 bins. The HET data have
slightly better spatial resolution than the PPAK observations, and the
PSF is given by the sum of two Gaussians with dispersions of 1\farcs19
and 3\farcs39, each weighted by 0.55 and 0.45, respectively.

The PPAK and HET kinematics extend out to $\sim$3 $R_e$ and show
features that are very similar to those seen from NIFS. The
large-scale kinematics reveal that the galaxy is rotating with
redshifted velocities of $\sim$160 km s$^{-1}$ to the northeast, a
peak in the central velocity dispersion to $\sim$350 km s$^{-1}$, and
a clear anti-correlation between $h_3$ and $V$. The PPAK and HET
kinematics are consistent with the NIFS kinematics over the radial
extent they share in common, after accounting for differences in
spatial resolution and binning.

\section{Stellar-Dynamical Models}
\label{sec:stellardyn_models}

In order to constrain the mass of the central black hole in Mrk 1216,
we calculated three-integral, orbit-based dynamical models using the
triaxial Schwarzschild code of \cite{vandenBosch_2008}. We ran the
code in the axisymmetric limit, meaning that triaxial orbit families
(e.g., box orbits) are included in the orbital libraries but we adopt
a nearly oblate axisymmetric shape with an intermediate to long axis
ratio of 0.99. The assumption of axisymmetry is justified by the
galaxy's rotation and the absence of isophotal and kinematic twists in
the \emph{HST} image and the NIFS/PPAK data. We deprojected the MGE in
Section \ref{sec:hst_img} using an inclination angle of
70$^\circ$. The same inclination was used by \cite{Yildirim_2015}, and
is mid-way between the range of angles for which the MGE can be
deprojected given the apparent axis ratio of the flattest Gaussian
component. Since Mrk 1216 does not contain a nuclear dust disk, we are
unable to derive an independent estimate of the inclination angle, as
has been possible for a few nearby galaxies
\citep[e.g.,][]{vandenBosch_2012, Yildirim_2016a}.

The stellar potential is combined with the gravitational potential due
to a black hole and a Navarro-Frenk-White \citep[NFW;][]{Navarro_1996}
dark matter halo. We created an orbit library that samples 32
equipotential shells with logarithmically spaced radii beginning at
0\farcs003, with 9 angular and 9 radial values at each energy. We
ensure a smooth distribution function by bundling together 125 orbits
with similar initial conditions. The 972,000 orbits were then
integrated in the galaxy's potential. We used a non-negative least
squares solver to assign weights to the orbits such that the
superposition matches the stellar kinematics, as well as the intrinsic
and projected stellar masses to an accuracy of 1\%, while accounting
for PSF effects and aperture binning.

The free parameters in the model are the black hole mass, the $H$-band
stellar mass-to-light ratio ($\Upsilon_H$), the concentration ($c$) of
the NFW halo, and the fraction of dark matter relative to the stellar
mass ($f_\mathrm{DM}$). The stellar mass-to-light ratio is assumed to
be constant with radius, which is supported by the lack of color
gradient in \emph{HST} WFC3 F814W and F160W images
\citep{Yildirim_2015}. We generated model grids that sampled 41 values
of $M_\mathrm{BH}$ with $8.5 \leq \log(M_\mathrm{BH}/M_\odot) \leq
10.5$, 28 values of $\Upsilon_H$ between 0.3 and $3.0\
\Upsilon_\odot$, and 24 NFW halos with $c=5,10,15$ and
$\log(f_\mathrm{DM})=0.5-4.0$. Models without a dark halo were run as
well. While the results of our fiducial model grid presented in
Section \ref{subsec:model_results} were obtained by fitting to the
NIFS$+$PPAK kinematics, in Section \ref{subsec:moremodels} we also
test fitting NIFS-only and NIFS+HET kinematics. With four
Gauss-Hermite moments in 108 NIFS$+$PPAK spatial bins, there are 432
observables. Ultimately, the best-fit model is the one with the lowest
$\chi^2$ ($\chi^2_\mathrm{min}$), and the 1$\sigma$ statistical
uncertainty for a given parameter is set by marginalizing over the
other free parameters and searching for where the change in $\chi^2$
($\Delta \chi^2 \equiv \chi^2 - \chi^2_\mathrm{min}$) is 1.0. The
3$\sigma$ model fitting error is taken to be where $\Delta \chi^2 =
9.0$.

\subsection{Modeling Results}
\label{subsec:model_results}

\begin{figure*}
\begin{center}
\epsscale{0.95}
\plottwo{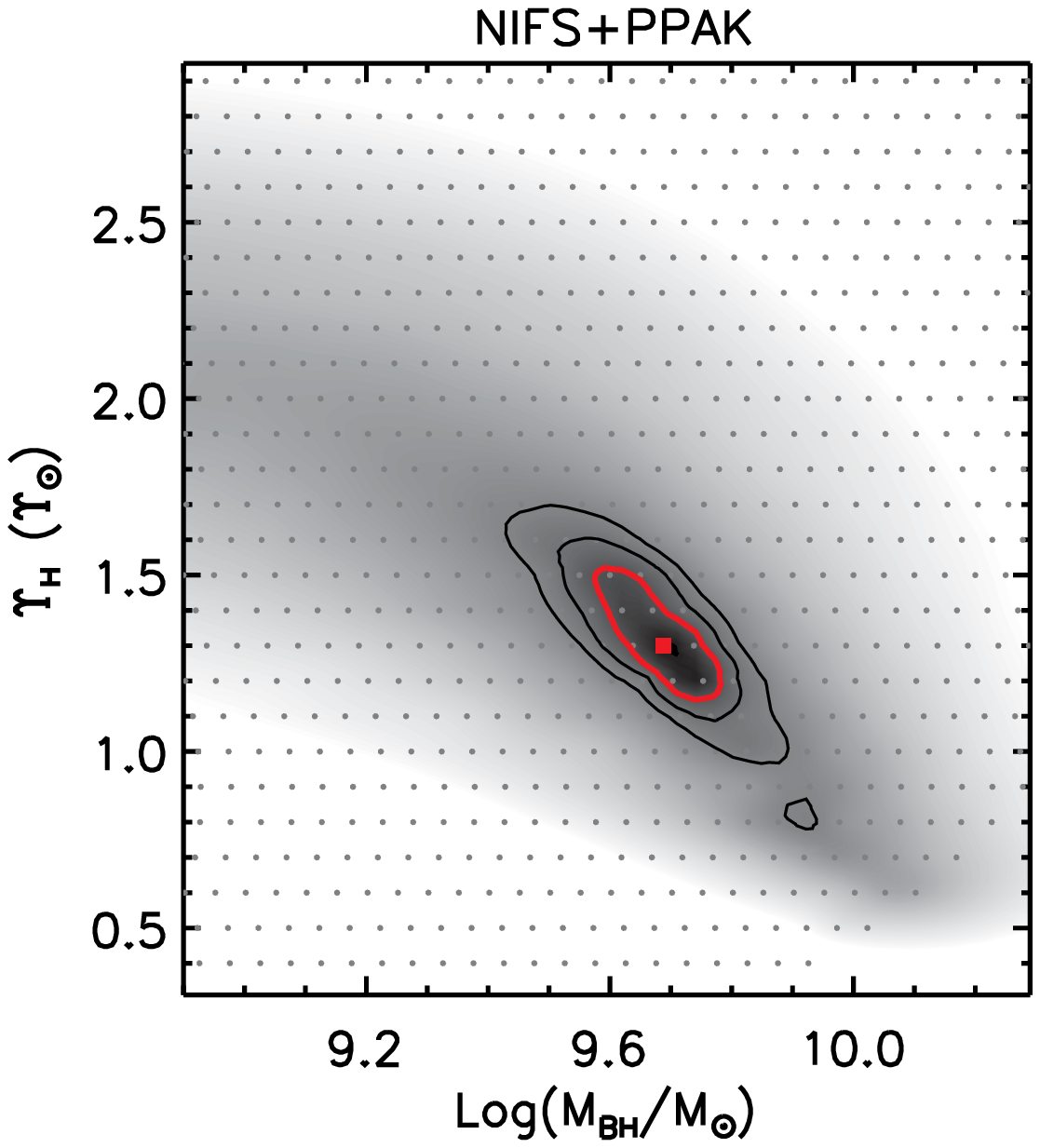}{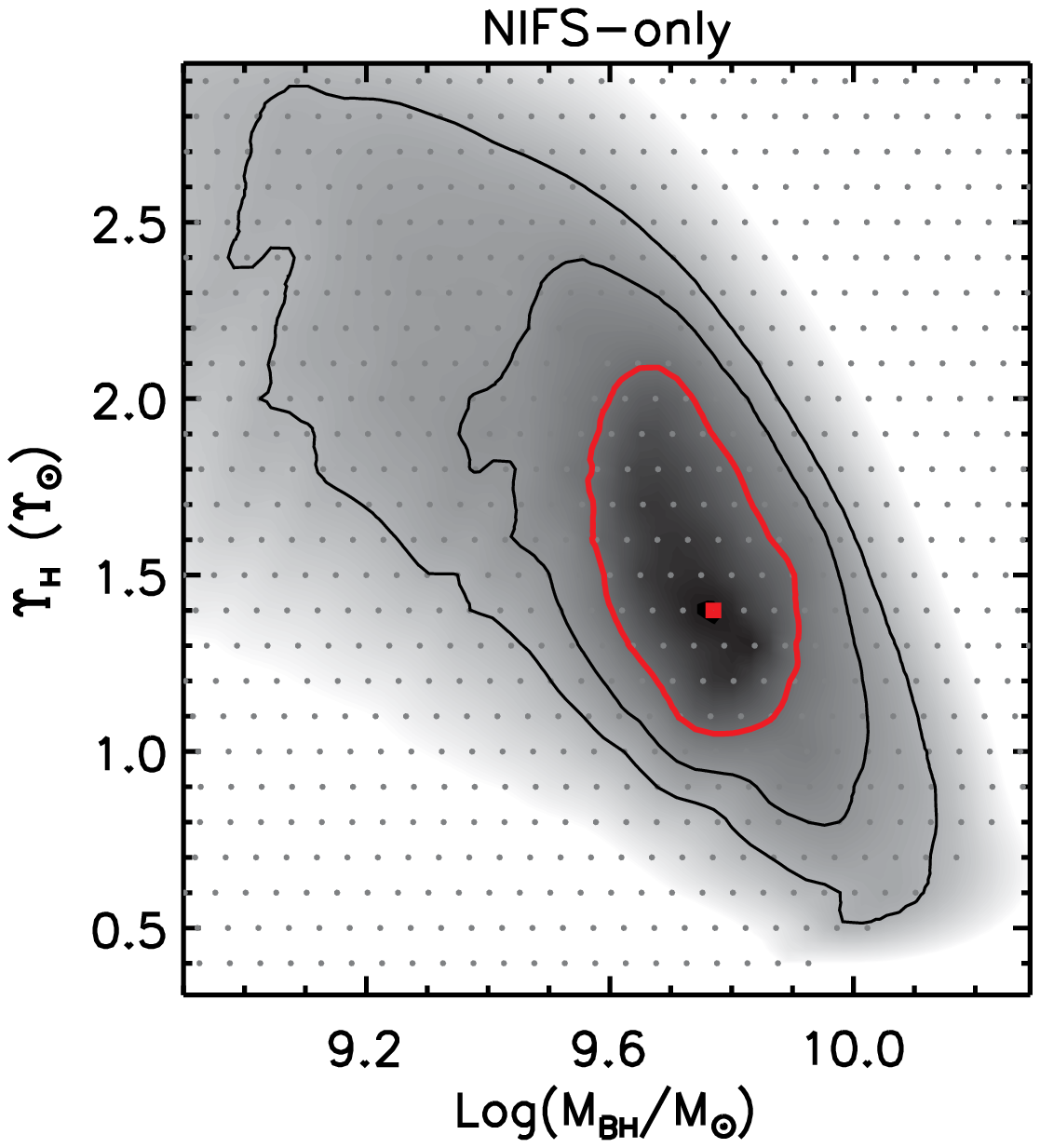}
\caption{Contours of $\chi^2$ for various stellar-dynamical models
  (gray dots) with different combinations of black hole mass and
  $H$-band stellar mass-to-light ratio after marginalizing over the
  dark halo parameters. The red square is the best-fit model, the red
  contour indicates where $\Delta \chi^2 = 2.3$, and the subsequent
  black contours correspond to $\Delta \chi^2 = 6.2$ and $11.8$,
  respectively. These $\Delta \chi^2 $ values correspond to 1$\sigma$,
  2$\sigma$, and 3$\sigma$ confidence regions for two parameters. The
  results are shown for dynamical models fit to the combination of
  NIFS and PPAK data sets (left) and for models fit to only the NIFS
  kinematics (right). The two grids produce consistent
  results. \label{fig:contours_nifsppak_nifs}}
\end{center}
\end{figure*}

\begin{figure*}
\begin{center}
\epsscale{0.85}
\plottwo{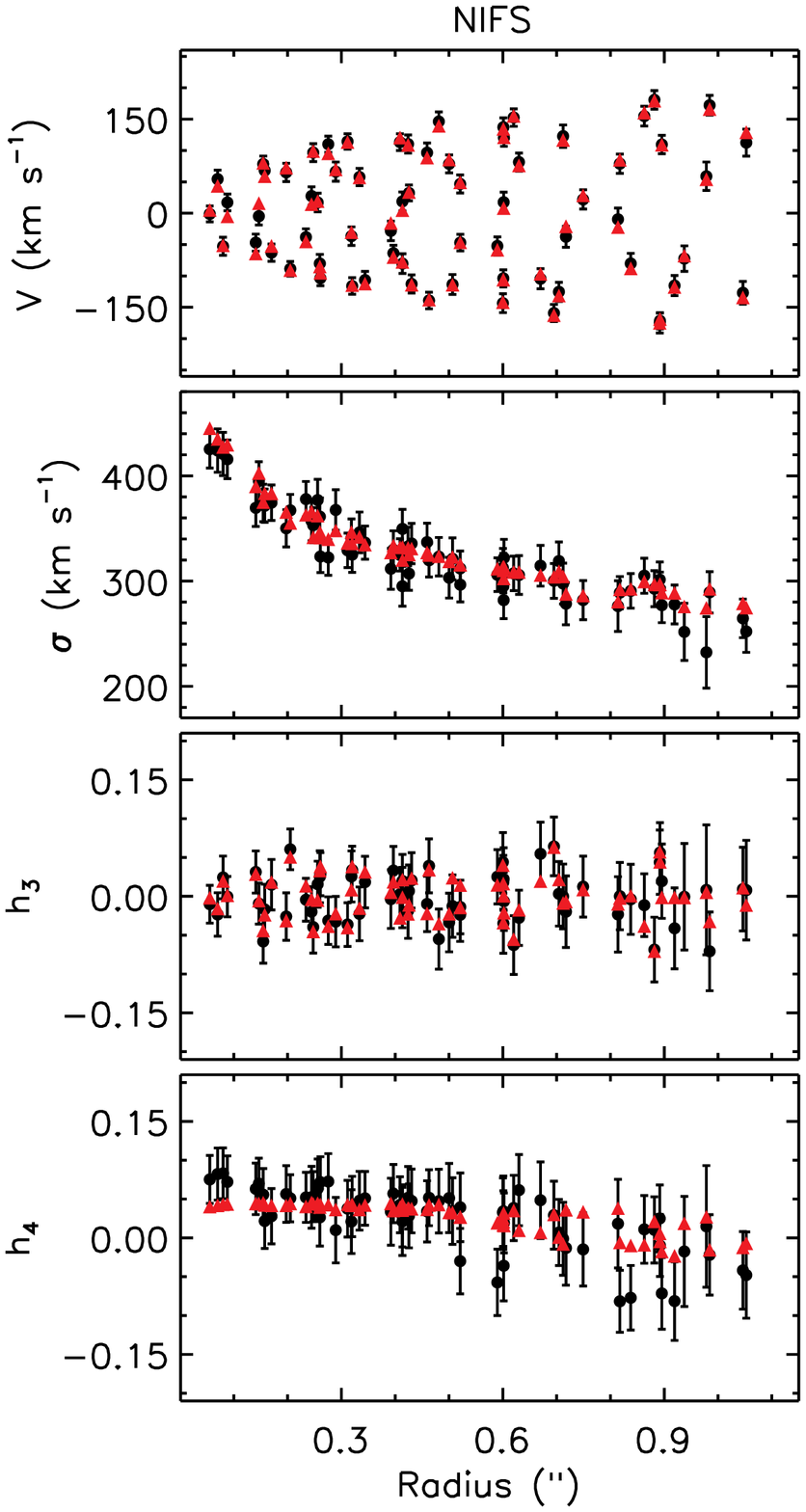}{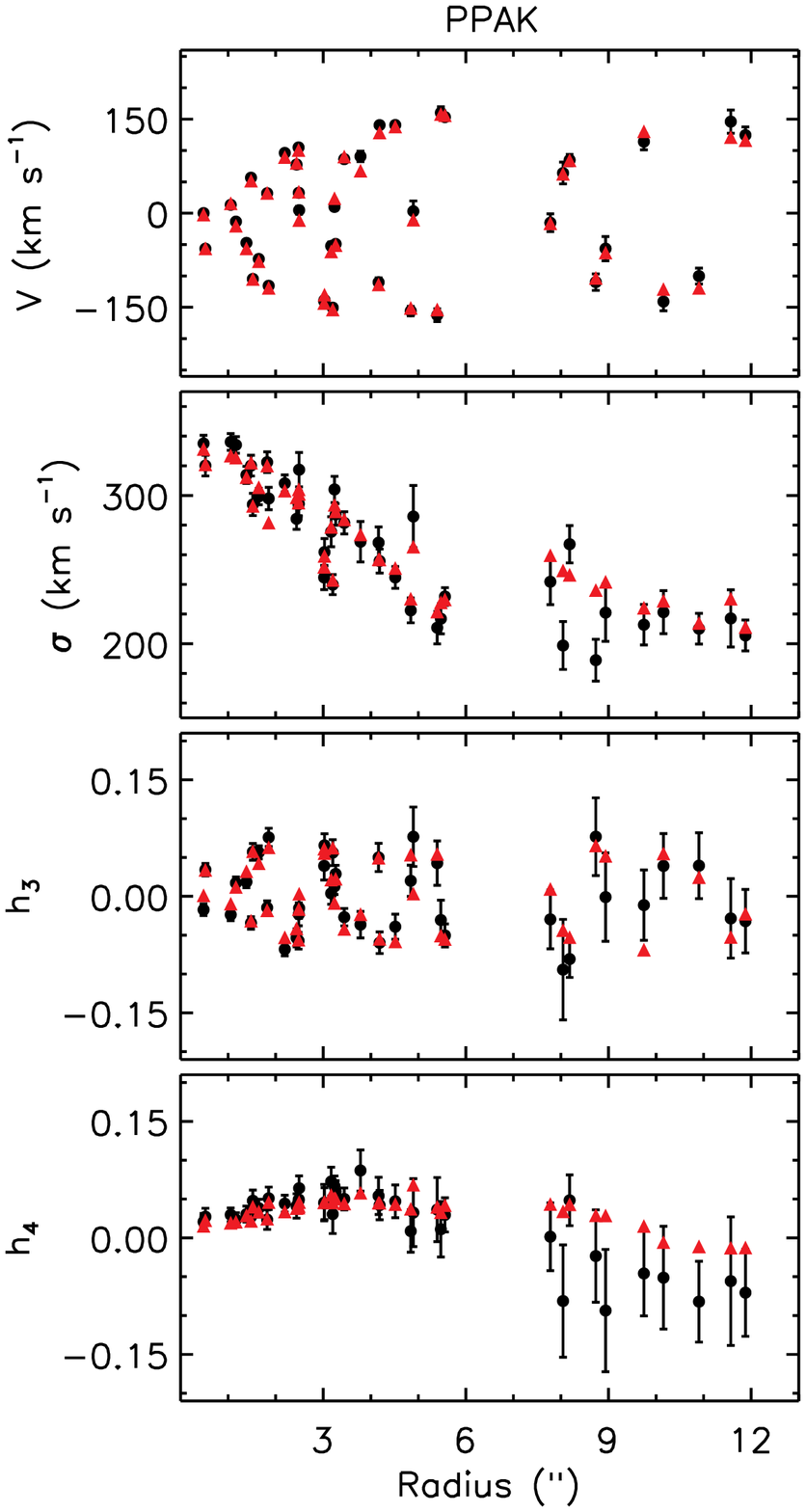}
\caption{The observed NIFS (left) and PPAK (right) kinematics, plotted
  as a function of projected radial distance from the nucleus, are
  compared to the best-fit stellar dynamical model (red) with
  $M_\mathrm{BH} = 4.9\times10^9\ M_\odot$ and $\Upsilon_H = 1.3\
  \Upsilon_\odot$. The data are folded and multiple position angles
  are depicted. The best-fit model reproduces the kinematic features
  well, and has a reduced $\chi^2$ of
  0.6. \label{fig:nifsppakmodel_nifsppak}}
\end{center}
\end{figure*}

We present the results of our fiducial model grid in the left panel of
Figure \ref{fig:contours_nifsppak_nifs}, and provide a comparison
between the best-fit model and NIFS/PPAK kinematics in Figure
\ref{fig:nifsppakmodel_nifsppak}. We find that $M_\mathrm{BH} =
4.9\times10^9\ M_\odot$, $\Upsilon_H = 1.3\ \Upsilon_\odot$, $c=10$,
and $\log(f_\mathrm{DM}) = 3.5$, which translates to a halo virial
mass of $5\times10^{14}\ M_\odot$. This model reproduces the observed
kinematics well, and has a reduced $\chi^2$ of 0.6. The 1$\sigma$
statistical uncertainties on $M_\mathrm{BH}$ and $\Upsilon_H$
correspond to $(4.9^{+0.8}_{-0.7})\times10^9\ M_\odot$ and $1.3\pm0.1\
\Upsilon_\odot$, respectively, whereas the 3$\sigma$ statistical
uncertainties translate to $M_\mathrm{BH} =
(4.9^{+1.8}_{-1.9})\times10^9\ M_\odot$ and $\Upsilon_H = 1.3\pm0.3\
\Upsilon_\odot$. In contrast, the dark halo parameters are not well
constrained. All three values of $c$ are allowed within 3$\sigma$ and
$\log(f_\mathrm{DM}) > 2.5$. Models without a dark halo are clearly
ruled out, as $\chi^2_\mathrm{min} = 320$ for the models without a
dark halo, which corresponds an increase of 65 relative to the
best-fit model incorporating an NFW dark matter halo.

Moreover, we build into the $M_\mathrm{BH}$ and $\Upsilon_H$ error
budgets the effect of adopting a different NIFS PSF model, using
unsymmetrized kinematics, and assuming a different inclination
angle. Due to the difficulties in measuring the AO PSF, it is
important to test how other reasonable PSF characterizations might
affect $M_\mathrm{BH}$. Our fiducial model above utilizes a
two-Gaussian description determined by comparing the galaxy's MGE to
the Mrk 1216 collapsed data cube. If instead the NIFS PSF is taken to
be the sum of three concentric, circular Gaussians measured from the
NIFS observation of the galaxy's tip-tilt star, we find
$M_\mathrm{BH} = 5.5\times10^9\ M_\odot$ and
$\Upsilon_H = 1.3\ \Upsilon_\odot$.

A similar change occurs when the observed stellar kinematics are not
forced to be point-symmetric. The fiducial model was fit to NIFS and
PPAK kinematics that were averaged in a two-fold manner over the major
and minor axes in order to reduce noise in the kinematic
measurements. When only the systematic offsets in the odd
Gauss-Hermite moments are subtracted off, but no other adjustments are
made, we find that $M_\mathrm{BH} = 5.8\times10^9\ M_\odot$ and
$\Upsilon_H = 1.1\ \Upsilon_\odot$. 

In addition to fitting to the point-symmetric NIFS$+$PPAK kinematics,
the fiducial model was run for an inclination angle of
70$^\circ$. Often Schwarzschild models are calculated for a single
viewing orientation, as it is computationally expensive to sample over
$M_\mathrm{BH}$, $\Upsilon_H$, two dark halo parameters, and the
inclination (or three angles in the case of triaxiality). In the few
cases where inclination was allowed to vary, the parameter was not
well constrained by the 2D line-of-sight kinematics
\citep{Krajnovic_2005, vandenBosch_vandeVen_2009,
  Walsh_2012}. Therefore, we determined the effect on $M_\mathrm{BH}$
and $\Upsilon_H$ if a near edge-on angle of $85^\circ$ is used
instead. We find that $M_\mathrm{BH}$ decreases by 22\% to
$3.8\times10^9\ M_\odot$ and $\Upsilon_H$ changes by 23\% to
$1.6\ \Upsilon_\odot$.

By adding in quadrature the 1$\sigma$ formal model fitting uncertainty
and the percent change in the best-fit values relative to the fiducial
model above, we ultimately find that $M_\mathrm{BH} =
(4.9\pm1.7)\times10^9\ M_\odot$ and $\Upsilon_H = 1.3\pm0.4\
\Upsilon_\odot$ for Mrk 1216. These values are consistent with models
fit to only the (bi-symmetrized) PPAK kinematics --
\cite{Yildirim_2015} found a black hole mass upper-limit of
$1.0\times10^{10}\ M_\odot$ and $\Upsilon_H = 1.8^{+0.5}_{-0.8}\
\Upsilon_\odot$, along with an NFW halo parameterized by $c=10$ and
$\log(f_\mathrm{DM}) = 2.9^{+1.1}_{-2.2}$ (3$\sigma$ statistical
uncertainties). In addition, our dynamical $H$-band mass-to-light
ratio is in agreement with expectations from stellar population
synthesis models for both a Kroupa ($1.2\ \Upsilon_\odot$) and a
Salpeter ($1.7\ \Upsilon_\odot$) initial mass function, assuming solar
metallicity and a $\sim$13 Gyr age \citep{Vazdekis_1996}. Given the
black hole mass of $4.9\times10^9\ M_\odot$ and the bulge stellar
velocity dispersion of 308 km s$^{-1}$ (see Section
\ref{sec:bhgalrels}), the NIFS data, with a central 0\farcs1 spatial
bin, have easily resolved the 0\farcs49 black hole sphere of
influence.

\subsection{Additional Models}
\label{subsec:moremodels}

The PPAK data cube provides an increase in 2D spatial coverage, more
spatial bins, smaller uncertainties on the extracted kinematics, and
better spectral resolution than the HET long-slit spectroscopy. Thus,
we use the PPAK kinematics in place of the HET kinematics when
constructing the dynamical models. If instead Schwarzschild models are
fit to the NIFS$+$HET kinematics, we recover the same results, with
$M_\mathrm{BH} = (4.9^{+0.8}_{-0.7})\times10^9\ M_\odot$ and
$\Upsilon_H = 1.3\pm0.1\ \Upsilon_\odot$ (1$\sigma$).

Likewise, we don't see significant changes in $M_\mathrm{BH}$ and
$\Upsilon_H$ if models are constrained by only the NIFS kinematics. By
fitting to the NIFS data alone, the black hole mass is less
susceptible to systematic effects that commonly plague
stellar-dynamical models, such as assumptions about the dark matter
halo \citep[e.g.,][]{Gebhardt_Thomas_2009, Schulze_Gebhardt_2011,
  Rusli_2013} and the radial form of the stellar mass-to-light ratio
\citep[e.g.,][]{McConnell_2013}. This comes at the expense of larger
$M_\mathrm{BH}$ statistical uncertainties due to the poor constraint
on $\Upsilon_H$. Results of fitting orbit-based models to the NIFS
kinematics alone are shown in the right panel of Figure
\ref{fig:contours_nifsppak_nifs}, and we find that $M_\mathrm{BH} =
(5.9^{+1.0}_{-1.7})\times10^9\ M_\odot$ and $\Upsilon_H =
1.4^{+0.5}_{-0.2}\ \Upsilon_\odot$ (1$\sigma$).

Finally, the NIFS kinematics were measured using an additive constant
and a degree 1 multiplicative polynomial to account for differences in
continuum shape between the velocity template library and the observed
galaxy spectra. We consider these NIFS kinematics robust, as changes
to how pPXF is run (see Section \ref{subsec:nifs_kinematics}) produce
similar kinematics. However, slight adjustments to the degree of the
additive/multiplicative polynomials can cause inconsistent kinematics
at the 1$\sigma$ level (all are consistent within 2$\sigma$). In
particular, one of the largest differences is seen when running pPXF
with a multiplicative degree 2 polynomial (with no additive
component), which produces 11 spatial bins in which the kinematics
differ by more than 1$\sigma$ relative to our adopted set of NIFS
kinematics. If instead we use the NIFS kinematics extracted with pPXF
and a multiplicative degree 2 polynomial, we infer $M_\mathrm{BH} =
(3.5^{+0.4}_{-0.5})\times10^9\ M_\odot$ and $\Upsilon_H = 1.3\pm0.1\
\Upsilon_\odot$ (1$\sigma$). We note that this second set of NIFS
kinematics show systematically smaller dispersions compared to the
central PPAK kinematics, hence we perform this test as a sanity check
but do not incorporate the results into our $M_\mathrm{BH}$ and
$\Upsilon_H$ error budgets.

The results from each of the three model grids above are in agreement
with our final black hole mass and mass-to-light ratio measurements
for Mrk 1216 of $M_\mathrm{BH} = (4.9\pm1.7)\times10^9\ M_\odot$ and
$\Upsilon_H = 1.3\pm0.4\ \Upsilon_\odot$.

\section{The Orbital Structure of Mrk 1216}
\label{sec:orbitstructure}

\begin{figure}
\begin{center}
\epsscale{1.05}
\plotone{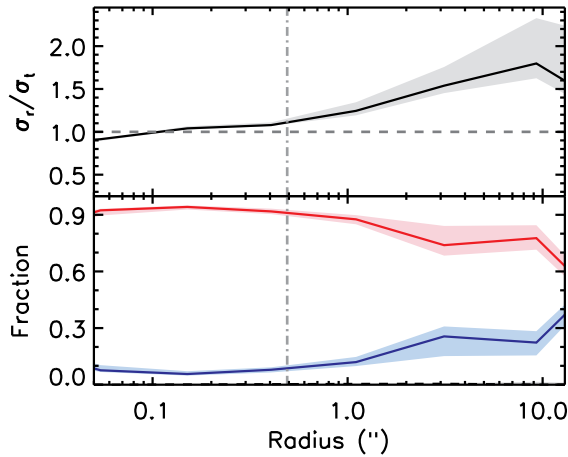}
\caption{Mrk 1216's orbital structure over the radial extent of the
  NIFS and PPAK kinematics. The anisotropy (top) and orbit type
  (bottom) are determined from the best-fit Schwarzschild model (solid
  lines) and a rough estimate of the uncertainties is depicted by the
  shaded regions. The galaxy is essentially isotropic (gray dashed
  horizontal line) within the black hole sphere of influence (gray
  dot-dashed vertical line), and becomes radially anisotropic at
  larger distances from the nucleus. The orbits are composed mainly of
  short-axis tubes (red), although a small fraction of box orbits
  (blue) are also present. \label{fig:anisotropy}}
\end{center}
\end{figure}

Not only do the Schwarzschild models provide us with constraints on
the black hole mass and the stellar mass-to-light ratio, but they also
allow for an examination of the galaxy's orbital structure. Figure
\ref{fig:anisotropy} illustrates the amount of anisotropy and the
orbit type as a function of radius. Using the best-fit model from
Section \ref{subsec:model_results}, we plot the ratio
$\sigma_r$/$\sigma_t$, where the tangential velocity dispersion is
given by $\sigma_t^2 = (\sigma_\phi^2 + \sigma_\theta^2)/2$ and $(r,
\theta, \phi)$ are spherical coordinates. In order to gain an idea of
the uncertainties in $\sigma_r$/$\sigma_t$, we show all the models
within $\Delta \chi^2 = 1$ from our fiducial grid, and the best-fit
models from the grid searches in which we assumed a different NIFS
PSF, unsymmetrized kinematics, and a near edge-on inclination
angle. We find that Mrk 1216 is roughly isotropic within the black
hole sphere of influence, but becomes radially anisotropic with
$\sigma_r$/$\sigma_t \sim 1.5$ at the radial extent of the PPAK
kinematics. The galaxy was modeled using a triaxial Schwarzschild code
\citep{vandenBosch_2008} that was run in the axisymmetric limit. Thus,
the best-fit model includes contributions from box orbits, albeit a
small one, making up 10\% of the orbits near the nucleus and 20\% at a
radius of 10\arcsec. Instead, short-axis tube orbits dominate our
best-fit model, contributing $\sim$$60-90$\% at all radii. Long-axis
tube orbits, which are important for prolate and triaxial systems, are
negligible.

\section{The Black Hole -- Host Galaxy Relations}
\label{sec:bhgalrels}

Placing Mrk 1216 on the black hole--host galaxy relations further
requires identifying the bulge component, and currently there is a
broad range of measurements in the literature. From Galfit models of
the \emph{HST} F160W image, \cite{Yildirim_2015} find an upper limit
on the $H$-band bulge luminosity and effective radius of $L_{H,
  \mathrm{bul}} = 8.3\times10^{10}\ L_\odot$ and $R_\mathrm{e, bul} =$
3\farcs42, and a lower limit of $L_{H, \mathrm{bul}} =
1.5\times10^{10}\ L_\odot$ and $R_\mathrm{e, bul} = $1\farcs34. These
numbers were derived from 2-component and 4-component 2D Sersic fits
for the upper and lower limits, respectively, and correspond to a
bulge-to-total ratio of $B/T = 0.69$ and $0.13$. The former
characterization includes a centrally concentrated component with a
Sersic index of 3.61 and a projected axis ratio of 0.56. Since this
double Sersic model produced pronounced residuals,
\cite{Yildirim_2015} increased the number of Sersic functions to four
to obtain a good fit to the galaxy's complex structure. Each of the
four components, however, have rather low Sersic indices between 0.99
and 1.61, which complicates a morphological classification. A
dynamical decomposition using orbital weights from the
\cite{Yildirim_2015} best-fit Schwarzschild model support the picture
depicted by the 4-component photometric decomposition. In contrast,
\cite{Savorgnan_Graham_2016} argue that Mrk 1216's bulge component has
an $H$-band luminosity of $\sim$$1.2\times10^{11}\ L_\odot$ based on a
one-dimensional (1D) multi-component Sersic fit to the \emph{HST}
F160W surface brightness brightness profile. Their fit includes an
intermediate-scale disk, in addition to a nuclear exponential disk and
a spheroidal component. Both \cite{Yildirim_2015} and
\cite{Savorgnan_Graham_2016} find that more than one Sersic component
is required to match the surface brightness despite the elliptical
galaxy classification in the NASA/IPAC Extragalactic Database and
Hyperleda.

For Mrk 1216, we conservatively adopt limits that extend from the
total quantities down to the smallest bulge measurements in the
literature, with values set to the result from a 2-component 2D Sersic
fit \citep{Yildirim_2015}. The galaxy's total luminosity was
determined from the MGE model in Section \ref{sec:hst_img}, which also
agrees with a single Sersic Galfit model of the \emph{HST} F160W
image. After assuming $H-K=0.2$ and a $K$-band absolute solar
magnitude of $3.28$ \citep{Binney_Merrifield_1998}, we establish a
bulge luminosity for Mrk 1216 of $L_{K, \mathrm{bul}} =
(9.6^{+4.4}_{-7.9})\times10^{10}\ L_\odot$. For comparison, the upper
bound on $L_{K, \mathrm{bul}}$ derived in this manner is 36\% smaller
than the growth curve analysis of Two Micron All Sky Survey
\citep{Skrutskie_2006} images by \cite{vandenBosch_2016}. We determine
the bulge mass using $\Upsilon_H = 1.3\ \Upsilon_\odot$ from the
best-fit model in Section \ref{subsec:model_results}. Although a
similar compact, high-dispersion galaxy, NGC 1277, showed evidence for
a radially varying $V$-band stellar mass-to-light ratio
\citep[e.g.,][]{MartinNavarro_2015}, there is no \emph{HST} WFC3
F814W$-$F160W color gradient observed for Mrk 1216
\citep{Yildirim_2015} and a similar $\Upsilon_H$ from dynamical models
fit to only the small-scale NIFS data in Section
\ref{subsec:moremodels} and models fit to only the large-scale PPAK
data \citep{Yildirim_2015} are found. Hence, adopting $\Upsilon_H =
1.3\ \Upsilon_\odot$ is justified and we find that $M_\mathrm{bul} =
(1.1^{+0.5}_{-0.9})\times10^{11}\ M_\odot$ for Mrk 1216. Finally, we
calculate $\sigma_\star$ for bulge effective radii of
$R_\mathrm{e,bul} = $1\farcs34, 3\farcs42, and 6\farcs34, which
correspond to the measurements from a 4-component, a 2-component, and
a 1-component Sersic Galfit model \citep{Yildirim_2015}. We use our
best-fit stellar-dynamical model to predict the luminosity-weighted
second moment within a circular aperture whose radius equals
$R_\mathrm{e,bul}$, while also excluding the region within the black
hole sphere of influence \citep[e.g.,][]{Gebhardt_2011,
  McConnell_Ma_2013}. Thus, we determine that $\sigma_\star =
308^{+16}_{-6}$ km s$^{-1}$ for Mrk 1216, which agrees well with the
previous measurement of the PPAK velocity dispersion within a circular
aperture that contains half of the light by \cite{Yildirim_2015}.

\begin{figure}
\begin{center}
\epsscale{1.14}
\plotone{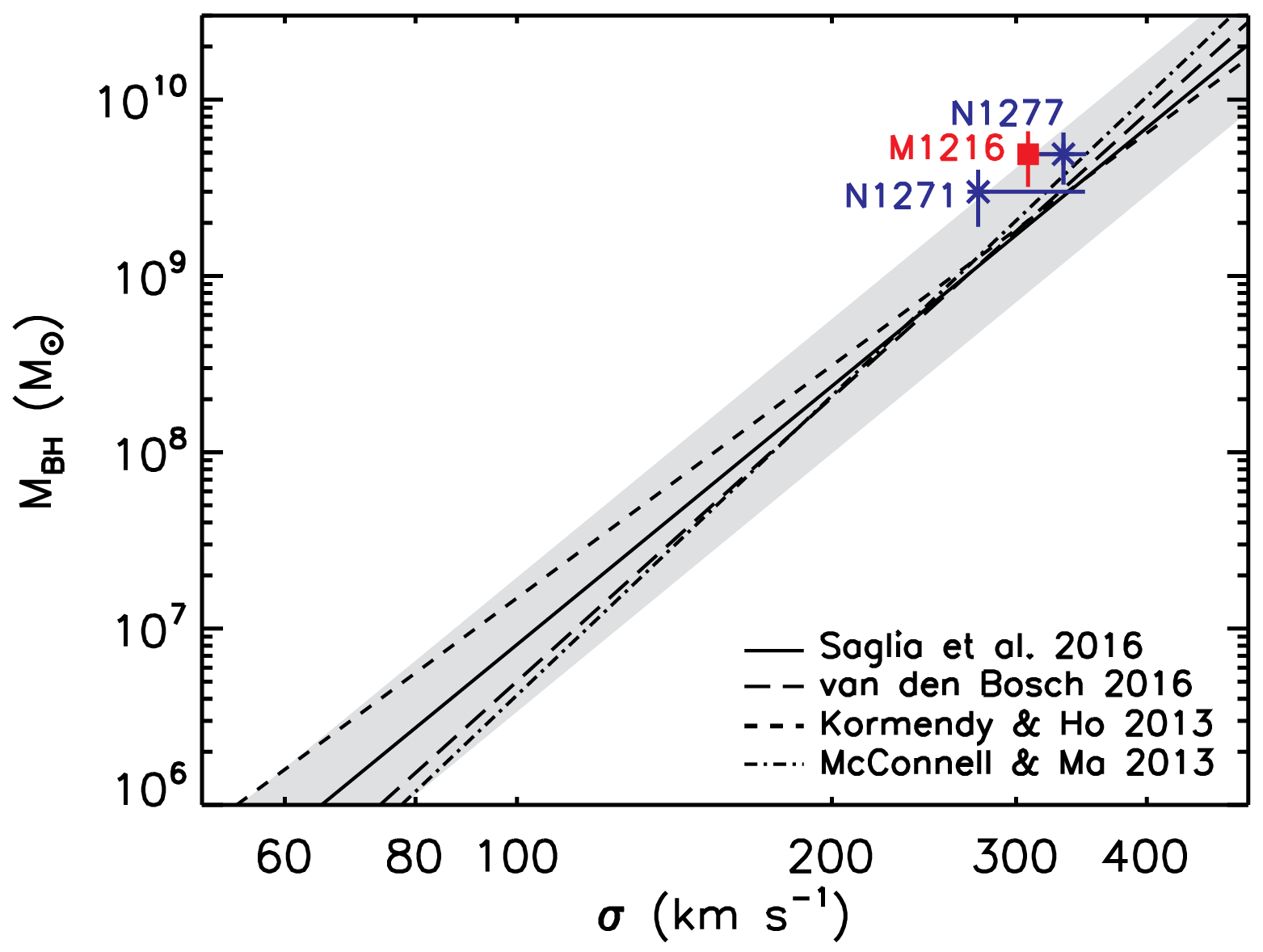}
\plotone{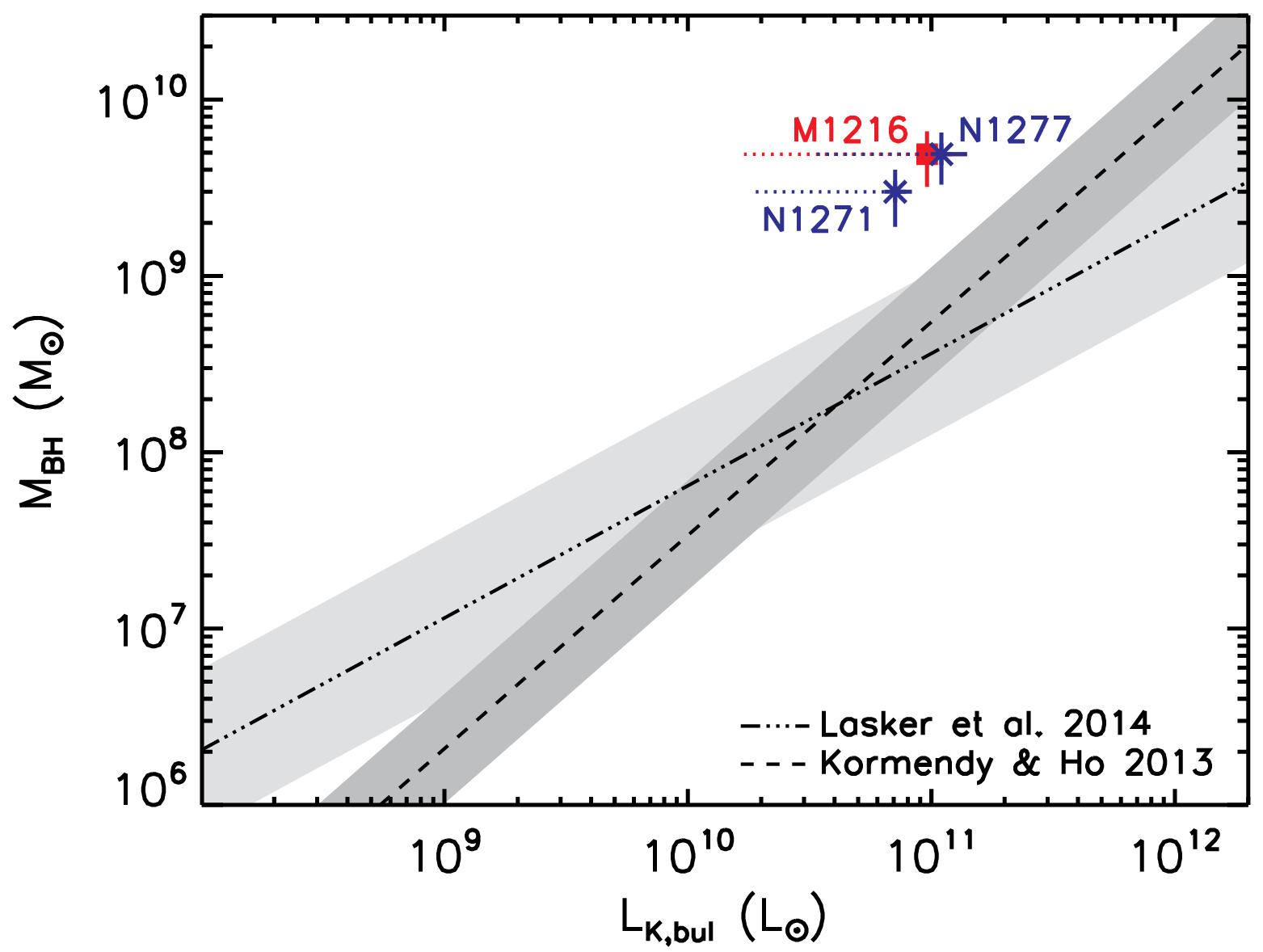}
\plotone{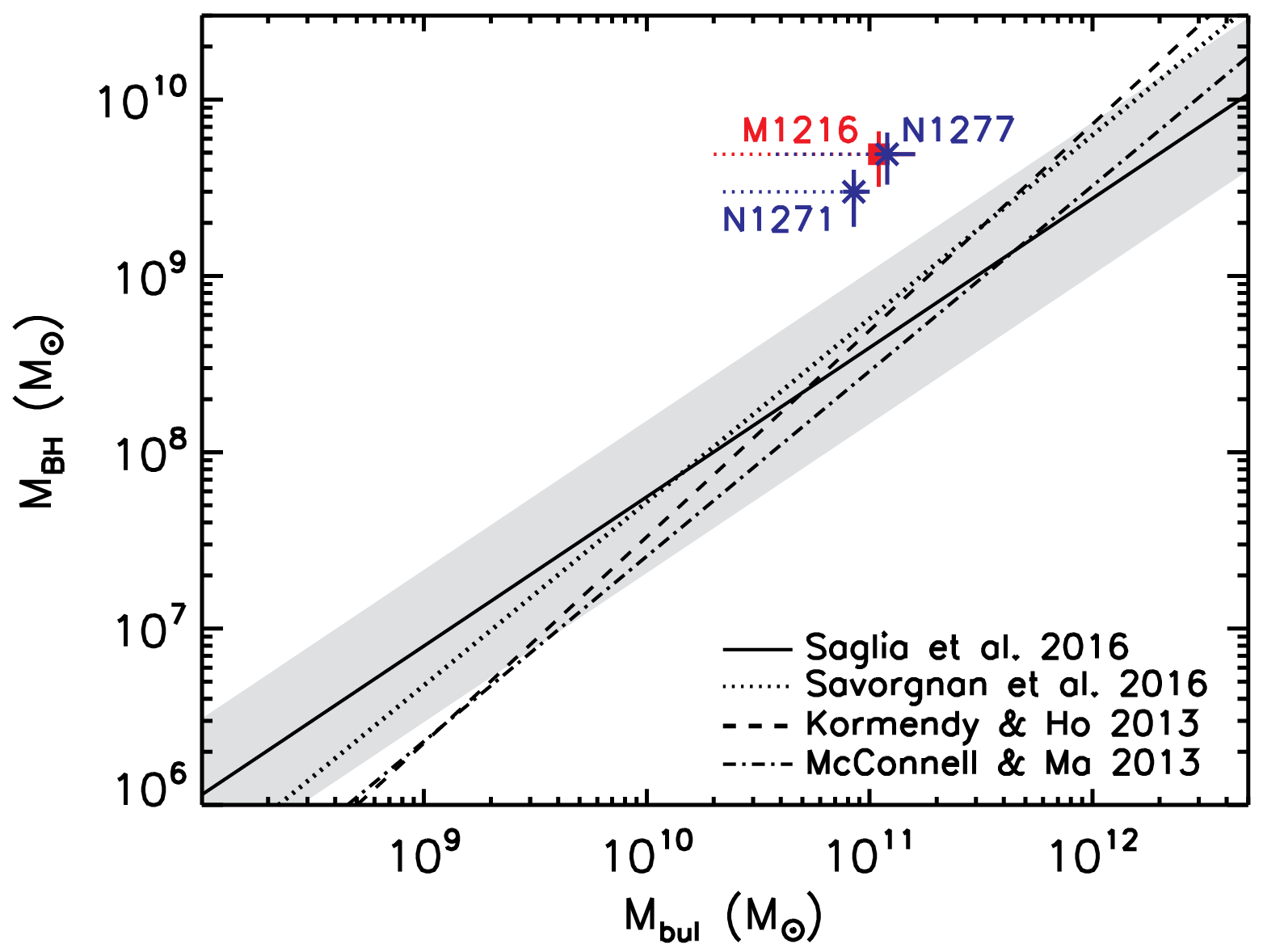}
\caption{Location of Mrk 1216 (red square) on the black hole -- host
  galaxy relations. We show multiple calibrations of the correlations,
  but for clarity only display the intrinsic scatter (gray) measured
  by \cite{Saglia_2016} for the black hole mass -- stellar velocity
  dispersion (top) and the black hole mass -- bulge mass (bottom)
  relationships. The intrinsic scatter from both
  \cite{Kormendy_Ho_2013} and \cite{Lasker_2014} are shown for the
  black hole mass -- $K$-band bulge luminosity (middle)
  relationship. NGC 1277 and NGC 1271 (blue asterisks) are two compact
  galaxies similar to Mrk 1216 from the HET Massive Galaxy Survey with
  $M_\mathrm{BH}$ measurements. Due to uncertainties in the bulge
  components, we show limits that extend from the total quantities
  (upper bound of the horizontal solid line) to the smallest bulge
  estimates (lower bound of the horizontal dotted line) in the bottom
  two panels. Even when using the total luminosity/stellar mass, Mrk
  1216, NGC 1277, and NGC 1271 are outliers from $M_\mathrm{BH} -
  L_{K,\mathrm{bul}}$ and $M_\mathrm{BH} - M_\mathrm{bul}$, yet are
  consistent with $M_\mathrm{BH} - \sigma_\star$. \label{fig:mbhrels}}
\end{center}
\end{figure}

\begin{figure*}
\begin{center}
\epsscale{0.7}
\plotone{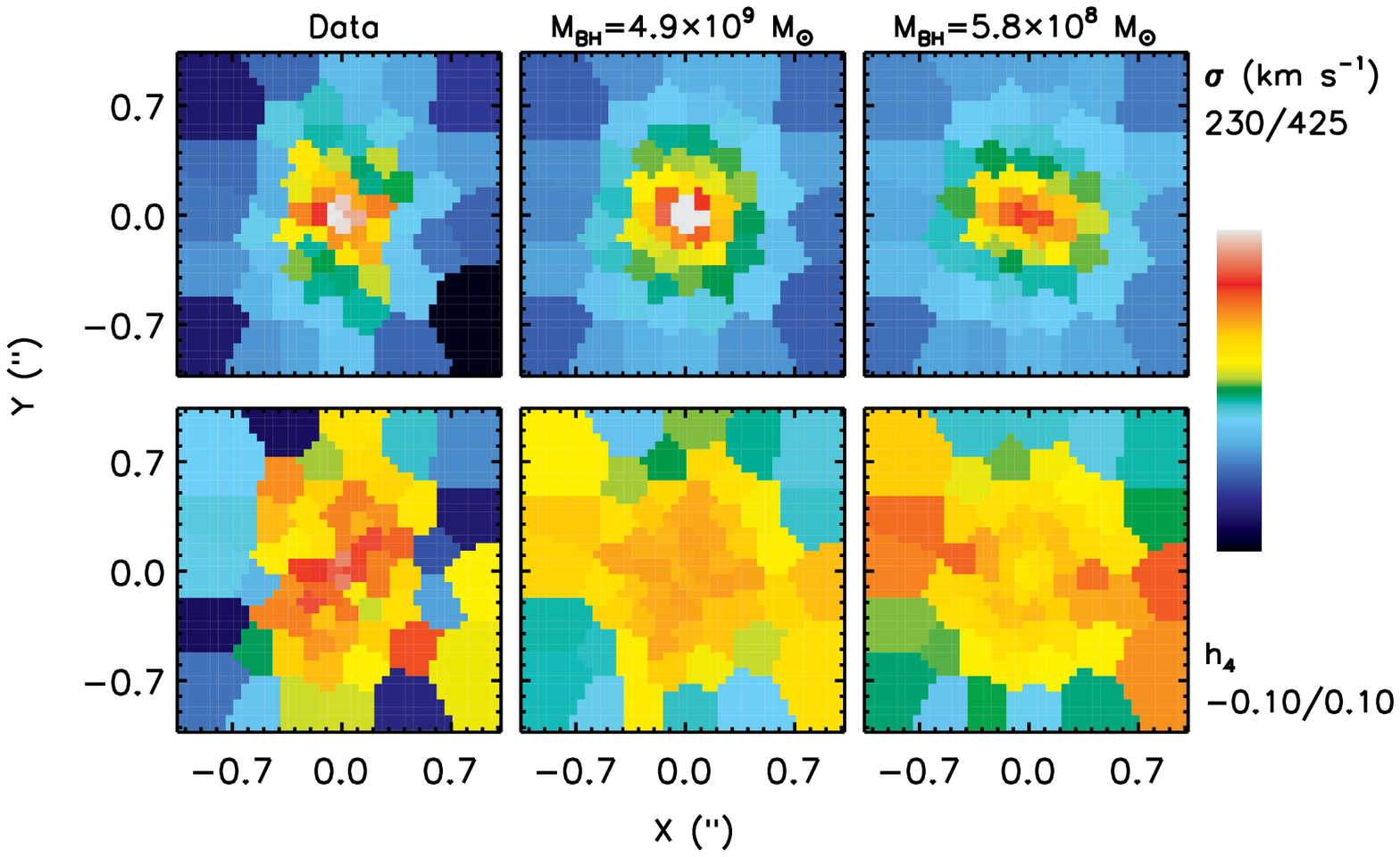}
\caption{Comparison of the stellar kinematics measured from NIFS
  (left) to the best-fit model with a $4.9\times10^9\ M_\odot$ black
  hole (middle) and a model with a $5.8\times10^8\ M_\odot$ black hole
  (right), which is the mass expected from
  $M_\mathrm{BH} - M_\mathrm{bul}$ \citep{Saglia_2016} when
  conservatively using Mrk 1216's total stellar mass of
  $1.6\times10^{11}\ M_\odot$. When generating models for the
  $5.8\times10^8\ M_\odot$ black hole, we sampled over a range stellar
  mass-to-light ratios and dark matter halos and present the model
  with the lowest $\chi^2$. The best-fit model is a good match to the
  data, while the model with a smaller black hole cannot reproduce the
  sharp rise in the velocity dispersion (top) or the elevated $h_4$
  values (bottom) at the nucleus. \label{fig:mbhmbulcompare}}
\end{center}
\end{figure*}

As can be seen in Figure \ref{fig:mbhrels}, Mrk 1216 is an outlier
from the $M_\mathrm{BH} - L_{K,\mathrm{bul}}$ and $M_\mathrm{BH} -
M_\mathrm{bul}$ relations, but is consistent with $M_\mathrm{BH} -
\sigma_\star$. Even when using the galaxy's total luminosity or total
stellar mass, Mrk 1216 falls a factor of $\sim$6 and $\sim$10 above
the \cite{Kormendy_Ho_2013} and \cite{Lasker_2014} $M_\mathrm{BH} -
L_{K,\mathrm{bul}}$ correlations, as well as a factor of $\sim$$5 -
10$ above $M_\mathrm{BH} - M_\mathrm{bul}$ depending on whether the
relation from \cite{McConnell_Ma_2013}, \cite{Kormendy_Ho_2013},
\cite{Savorgnan_2016}, \cite{Saglia_2016} is assumed. Thus, using the
total luminosity (stellar mass) makes Mrk 1216 a $2.2-2.5\sigma$
outlier ($1.4-3.0\sigma$ outlier) from the $M_\mathrm{BH} -
L_{K,\mathrm{bul}}$ ($M_\mathrm{BH} - M_\mathrm{bul}$) relation given
the various calibrations and scatter of the correlations. In Figure
\ref{fig:mbhmbulcompare}, we show predictions of a model with a
$5.8\times10^8\ M_\odot$ black hole, which is the mass expected from
$M_\mathrm{BH} - M_\mathrm{bul}$ \citep{Saglia_2016} for Mrk 1216's
total stellar mass of $1.6\times10^{11}\ M_\odot$. These $\sigma$ and
$h_4$ predictions are compared to the NIFS observations and the
best-fit model from Section \ref{subsec:model_results} with
$M_\mathrm{BH} = 4.9\times10^9\ M_\odot$. Our best-fit model exhibits
a similar central velocity dispersion peak and elevated central $h_4$
values as the observed kinematics, whereas the model based on
$M_\mathrm{BH} - M_\mathrm{bul}$ cannot reproduce these features.

For comparison, the two other compact, high-dispersion galaxies from
the HET Massive Galaxy Survey are shown in Figure
\ref{fig:mbhrels}. We follow the same conventions for characterizing
the NGC 1277 and NGC 1271 bulge quantities as was used for Mrk
1216. In particular, from the MGE descriptions of \emph{HST} images
\citep{vandenBosch_2012, Walsh_2015} and the best-fit stellar
mass-to-light ratios from dynamical models fit to AO observations
\citep{Walsh_2015, Walsh_2016}, we measure total luminosities of $L_V
= 1.7\times10^{10}\ L_\odot$ and $L_H = 7.2\times10^{10}\ L_\odot$,
and total stellar masses of $1.6\times10^{11}\ M_\odot$ and
$1.0\times10^{11}\ M_\odot$ for NGC 1277 and NGC 1271,
respectively. These values are similar to the total luminosities and
masses reported by \cite{Emsellem_2013} for NGC 1277 and
\cite{Graham_2016b} for NGC 1271. \cite{Graham_2016a} calculate a
larger total luminosity for NGC 1277 based on modeling the 1D light
profile, however their MGE model suggests a smaller total luminosity
than the one adopted here (their MGE model has 43\% less light than
their 1D component analysis). Despite using total properties, NGC 1277
and NGC 1271 remain outliers from $M_\mathrm{BH} - L_{K,
  \mathrm{bul}}$ by $2.1 - 2.8\sigma$ \citep{Kormendy_Ho_2013,
  Lasker_2014} and $M_\mathrm{BH} - M_\mathrm{bul}$ by $1.4 -
3.0\sigma$ \citep{McConnell_Ma_2013, Kormendy_Ho_2013, Savorgnan_2016,
  Saglia_2016}. The two galaxies are in good agreement with the
expectations from $M_\mathrm{BH} - \sigma_\star$
\citep{McConnell_Ma_2013, Kormendy_Ho_2013, Saglia_2016,
  vandenBosch_2016}.

\cite{vandenBosch_2016} conclude that $M_\mathrm{BH} - \sigma_\star$
is the best empirical relationship available and that a multi-variate
scaling relation between $M_\mathrm{BH}$, $L_K$ and $R_e$ is a
projection of $M_\mathrm{BH} - \sigma_\star$ with an equal amount of
intrinsic scatter. Previous attempts to explore whether the inclusion
of an additional parameter leads to tighter scaling relations
\citep[e.g.,][]{Beifiori_2012, Saglia_2016} have also generally found
no significant decreases in the amount of intrinsic scatter compared
to the single-parameter $M_\mathrm{BH} - \sigma_\star$ relation. Given
the arguments of \cite{vandenBosch_2016}, it is not surprising that
these compact galaxies are consistent with $M_\mathrm{BH} -
\sigma_\star$ but with their small sizes are outliers from
$M_\mathrm{BH} - L_\mathrm{bul}$. The three HET compact galaxies are
in the compilation of \cite{vandenBosch_2016}. With this new black
hole mass and luminosities derived from the \emph{HST} images, we find
that Mrk 1216 and NGC 1271, are within $\sim$$1.7\sigma$ of the black
hole mass -- galaxy luminosity -- galaxy size relation correlation,
and that NGC 1277 is consistent with the relation, given the intrinsic
scatter. Future work on multi-variate scaling relations requires
overcoming a sampling bias in the known $M_\mathrm{BH}$ hosts;
currently there is a very limited spread in effective radii at a given
$K$-band galaxy luminosity \citep{vandenBosch_2015}.

\section{Discussion}
\label{sec:discussion}

Mrk 1216 hosts one of the largest black holes dynamically detected to
date, naturally leading to the question of how such a massive black
hole ended up in a relatively modest galaxy. One interesting
explanation is that Mrk 1216 is a relic of the $z \sim 2$ quiescent
galaxies, and avoided the same series of mergers that produced the
typical massive early-type galaxies of today. Mrk 1216 perhaps
reflects an earlier period when galaxies contained over-massive black
holes for their bulge luminosities/masses, and galaxy growth had yet
to follow. Such a scenario has been proposed to explain the locations
of NGC 1277 and NGC 1271 on the black hole scaling relations based
upon having small effective radii for the stellar masses, stellar mass
surface density profiles comparable to the $z \sim 2$ red nuggets, and
a uniformly old stellar population out to several $R_e$ (e.g.,
\citealt{Trujillo_2014, FerreMateu_2015, MartinNavarro_2015};
Y{\i}ld{\i}r{\i}m {et~al.} 2016b, in prep). Mrk 1216 is akin to NGC
1277 and NGC 1271, and is similar to the $z \sim 2$ red nuggets. The
galaxy is a clear outlier from the local galaxy mass--size relation,
and has an elevated central stellar mass surface density
(Y{\i}ld{\i}r{\i}m {et~al.} 2016b, in prep). However, there are hints
that Mrk 1216 has experienced some growth and begun the process of
becoming a normal early-type galaxy. Mrk 1216 has one of the largest
effective radii ($R_e = 2.8$ kpc) in our sample of compact galaxies
found through the HET Massive Galaxy Survey, and the stellar mass
surface density profile is more extended in the outer regions
(Y{\i}ld{\i}r{\i}m {et~al.} 2016b, in prep). With future work, it
would be informative to study Mrk 1216's stellar population over the
extent of the galaxy.

Another possible explanation for the location of Mrk 1216 on the
$M_\mathrm{BH} - L_\mathrm{bul}$ and $M_\mathrm{BH} - M_\mathrm{bul}$
relations is that we simply do not have enough measurements at the
upper-end of the correlations. With the limited number of objects in
this high-mass regime, neither the form of the correlations nor the
magnitude and distribution of the scatter are well determined
\citep[e.g.,][]{McConnell_Ma_2013}. Therefore, Mrk 1216 could fall in
the tails of a distribution between black hole mass and galaxy
properties that still needs to be fully flushed out. We note that many
of the compact galaxies in the HET Massive Galaxy Survey have nuclear
dust disks, indicating the presence of cleanly rotating gas
\citep{Ho_2002, Alatalo_2013}, from which independent black hole mass
measurements can be derived for comparison to the stellar-dynamical
determinations. Such cross-checks between mass measurement methods is
essential for establishing the amount of intrinsic scatter in the
black hole correlations, and eventually for assessing how strongly Mrk
1216, NGC 1277, and NGC 1271 deviate from the relations. Presently, a
majority of the meaningful comparisons between the stellar and
gas-dynamical methods have led to discrepancies where the
stellar-dynamical $M_\mathrm{BH}$ exceeds the gas-dynamical mass by
factors of $2-3$ \citep{deFrancesco_2006, Rusli_2011, Gebhardt_2011,
  Walsh_2012, Walsh_2013, Barth_2016b}, although there are a very
small number of direct comparison studies.

Finally, over-massive black holes can result from tidal stripping
events. This idea was used to explain the low-mass galaxy M60-UCD1,
which lies a mere 6.6 kpc away from the giant elliptical galaxy M60
\citep{Seth_2014}. Indeed, the EAGLE cosmological, hydrodynamical
simulation \citep{Schaye_2015, Crain_2015} shows that tidal stripping
is the dominant process responsible for the extreme outliers from the
$M_\mathrm{BH} - M_\star$ correlation, but the simulation is most
sensitive to galaxies with $M_\mathrm{BH} \sim 10^8\ M_\odot$ and
$M_\star \sim 10^{10}\ M_\odot$. Due to the limited box size,
predictions cannot be made for more massive NGC 1277-like galaxies
\citep{Barber_2016}. Contrary to NGC 1277 and NGC 1271, which are
members of the Perseus cluster, Mrk 1216 is an isolated galaxy in the
field, with only two other galaxies within 1 Mpc at its distance
\citep{Yildirim_2015}. Combined with the regular isophotes and lack of
tidal signatures in the \emph{HST} image, the idea that Mrk 1216 was
once the center of a more massive galaxy seems unlikely.

\section{Conclusion}
\label{sec:conclusion}

In summary, we measured the 2D stellar kinematics of the compact,
high-dispersion galaxy Mrk 1216 using newly acquired AO NIFS
observations that probe within the black hole sphere of influence. Mrk
1216 is rotating, has a distinct rise in the stellar velocity
dispersion at the nucleus, exhibits the expected anti-correlation
between $V$ and $h_3$, and has elevated central $h_4$ values. The high
angular resolution NIFS kinematics, along with large-scale kinematic
measurements and the luminous mass model from an \emph{HST} image, are
fit with stellar-dynamical models based upon the Schwarzschild
superposition method. We constrain the mass of the central black hole
in Mrk 1216 to be $(4.9\pm1.7)\times10^9\ M_\odot$ and the $H$-band
stellar mass-to-light ratio to be $1.3\pm0.4\ \Upsilon_\odot$. The
error budget incorporates some possible systematic effects and the
formal 1$\sigma$ model fitting uncertainties.

With $\sigma_\star = 308$ km s$^{-1}$, Mrk 1216 is consistent with the
$M_\mathrm{BH} - \sigma_\star$ relationship, but is a surprising
positive outlier from $M_\mathrm{BH} - L_\mathrm{bul}$ and
$M_\mathrm{BH} - M_\mathrm{bul}$, even when conservatively using the
galaxy's total luminosity and stellar mass of $L_{K} =
1.4\times10^{11}\ L_\odot$ and $M_\star = 1.6\times10^{11}\
M_\odot$. Mrk 1216 is similar to NGC 1277 and NGC 1271 -- the two
compact, high-dispersion galaxies from the HET Massive Galaxy Survey
that have prior stellar-dynamical $M_\mathrm{BH}$ measurements from AO
observations. All three galaxies resemble the quiescent galaxies at $z
\sim 2$ given their small but massive nature, their stellar mass
surface density profiles, their strong rotation, and (in the cases of
NGC 1277 and NGC 1271) their uniformly old stellar
populations. Therefore, Mrk 1216, as well as NGC 1277 and NGC 1271,
appear to be relics of the $z \sim 2$ red nuggets, and their black
holes may imply that the normalization of the $M_\mathrm{BH} -
L_\mathrm{bul}$ and $M_\mathrm{BH} - M_\mathrm{bul}$ relations were
higher at earlier times. In other words, perhaps black hole growth
precedes that of its host galaxy. Another possibility is that the
galaxies are simply unusual and are in the tail of a distribution
between $M_\mathrm{BH}$ and galaxy properties that still needs to be
firmly established. Distinguishing between the two scenarios requires
obtaining a more complete census of local black holes in a wide range
of galaxies with diverse evolutionary histories.

\acknowledgements

Based on observations obtained at the Gemini Observatory acquired
through the Gemini Science Archive and processed using the Gemini IRAF
package, which is operated by the Association of Universities for
Research in Astronomy, Inc., under a cooperative agreement with the
NSF on behalf of the Gemini partnership: the National Science
Foundation (United States), the National Research Council (Canada),
CONICYT (Chile), the Australian Research Council (Australia),
Minist\'{e}rio da Ci\^{e}ncia, Tecnologia e Inova\c{c}\~{a}o (Brazil)
and Ministerio de Ciencia, Tecnolog\'{i}a e Innovaci\'{o}n Productiva
(Argentina), under program GN-2013A-Q-1. Also based on observations
made with the NASA/ESA Hubble Space Telescope, obtained at the Space
Telescope Science Institute, which is operated by the Association of
Universities for Research in Astronomy, Inc., under NASA contract NAS
5-26555. These observations are associated with program \#13050. Based
on observations collected at the Centro Astron\'{o}mico Hispano
Alem\'{a}n (CAHA) at Calar Alto, operated jointly by the Max-Planck
Institut f\"{u}r Astronomie and the Instituto de Astrof\'{i}sica de
Andaluc\'{i}a (CSIC). This work is further based on observations
obtained with the Hobby-Eberly Telescope (HET). The HET is a joint
project of the University of Texas at Austin, the Pennsylvania State
University, Ludwig-Maximilians-Universit\"{a}t M\"{u}nchen, and
Georg-August-Universit\"{a}t G\"{o}ttingen. The HET is named in honor
of its principal benefactors, William P. Hobby and Robert
E. Eberly. This material is based in part upon work J.~L.~W. conducted
as an NSF Astronomy and Astrophysics Postdoctoral Fellow under Award
No.~1102845. The authors further acknowledge the Texas Advanced
Computing Center (TACC; http://www.tacc.utexas. edu) at the University
of Texas at Austin for providing HPC resources that have contributed
to the research results reported within this paper. This research has
made use of the NASA/IPAC Extragalactic Database which is operated by
the Jet Propulsion Laboratory, California Institute of Technology,
under contract with NASA. We acknowledge the usage of the HyperLeda
database (http://leda.univ-lyon1.fr).

\end{document}